\documentclass[aps,prx,twocolumn,longbibliography]{revtex4-1}

\usepackage{siunitx}
\usepackage{amsmath}
\usepackage{amssymb}
\usepackage{graphicx}
\usepackage[colorlinks,allcolors= blue]{hyperref}
\usepackage{ulem}

\begin{document}
\title{Coherent feedback cooling of a nanomechanical membrane with atomic spins}

\author{Gian-Luca Schmid}
\altaffiliation{These authors contributed equally to this work}
\affiliation{Department of Physics and Swiss Nanoscience Institute, University of Basel, 4056 Basel, Switzerland}
\author{Chun Tat Ngai}
\altaffiliation{These authors contributed equally to this work}
\affiliation{Department of Physics and Swiss Nanoscience Institute, University of Basel, 4056 Basel, Switzerland}
\author{Maryse Ernzer}
\affiliation{Department of Physics and Swiss Nanoscience Institute, University of Basel, 4056 Basel, Switzerland}
\author{Manel Bosch Aguilera}
\affiliation{Department of Physics and Swiss Nanoscience Institute, University of Basel, 4056 Basel, Switzerland}
\author{Thomas M. Karg}
\altaffiliation{Current address: IBM Research Europe, Zurich, S\"aumerstrasse 4, CH-8803 R\"uschlikon, Switzerland}
\affiliation{Department of Physics and Swiss Nanoscience Institute, University of Basel, 4056 Basel, Switzerland}
\author{Philipp Treutlein}
\email{philipp.treutlein@unibas.ch}
\affiliation{Department of Physics and Swiss Nanoscience Institute, University of Basel, 4056 Basel, Switzerland}

\begin{abstract}
Coherent feedback stabilises a system towards a target state without the need of a measurement, thus avoiding the quantum backaction inherent to measurements.  Here, we employ optical coherent feedback to remotely cool a nanomechanical membrane using atomic spins as a controller. Direct manipulation of the atoms allows us to tune from strong-coupling to an overdamped regime. Making use of the full coherent control offered by our system, we perform spin-membrane state swaps combined with stroboscopic spin pumping to cool the membrane in a room-temperature environment to ${T}=\SI{216}{\milli\kelvin}$ ($\bar{n}_{m} = 2.3\times 10^3$ phonons) in $\SI{200}{\micro\second}$. We furthermore observe and study the effects of delayed feedback on the cooling performance. Starting from a cryogenically pre-cooled membrane, this method would enable cooling of the mechanical oscillator close to its quantum mechanical ground state and the preparation of nonclassical states.
\end{abstract}

\maketitle 

\section{Introduction}
Hybrid quantum systems in which a mechanical oscillator is coupled to a spin are a promising platform for fundamental quantum science as well as for quantum sensing \cite{treutlein2014, kurizki2015,chu2020}. 
The interest in such systems derives from the fact that the spin -- a genuinely quantum-mechanical object -- can be used to control, read-out, and lend new functionality to the much more macroscopic mechanical device. 
Recently, different spin-mechanics interfaces have been realized, involving the coupling of a mechanical oscillator to (pseudo-)spin systems such as atomic ensembles \cite{camerer2011,jockel2015, christoph2018,moller2017,karg2020,thomas2021}, quantum dots \cite{yeo2014,montinaro2014}, superconducting qubits  \cite{oconnell2010,arrangoiz-arriola2019,clerk2020}, or impurity spins in solids \cite{rugar2004,arcizet2011,barfuss2015,lee2017}, using light-, strain-, or magnetically-mediated interactions.

Coherent feedback is an intriguing concept that can be studied with such systems \cite{lloyd2000,zhang2017}. In coherent feedback, a quantum system is controlled through its interaction with another one, in such a way that quantum coherence is preserved. In contrast to measurement-based feedback \cite{wiseman2009}, coherent feedback does not rely on measurements, thus avoiding the associated backaction and decoherence. Coherent feedback can under certain conditions outperform measurement-based feedback in tasks such as cooling of resonators \cite{hamerly2012,bennett2014}, and it has been implemented in solid state systems to enhance the coherence time of a qubit \cite{hirose2016}. In optomechanical systems, it has been theoretically studied as a way to generate large nonlinearities at the single photon level  \cite{zhang2012,wang2017a}, to enhance optomechanical cooling and state transfer \cite{harwood2021}, as well as for entanglement generation \cite{woolley2014,li2017,harwood2021}. 

In the context of spin-mechanics interfaces, the mechanical oscillator can act as the system to be controlled, i.e.\ the \textit{plant}, which is coupled to a noisy thermal bath, and the spin system as the \textit{controller}, coupled to a zero-temperature bath. Coherent feedback is achieved by coupling the two systems, thus reducing the noise in the mechanical system by transferring it to the spin, where it is dissipated.  Additional coherent control of the spin enhances the cooling performance.

Hybrid systems combining atomic ensembles and mechanical oscillators have been used for sympathetic cooling by coupling the mechanical vibrations of a membrane to the center-of-mass oscillation of cold atoms in an optical lattice \cite{jockel2015,christoph2018}. In these systems the atomic motion was strongly damped and did not offer the possibility for coherent control. Furthermore, optical traps for atoms cannot reach MHz trapping frequencies without introducing substantial photon scattering and dissipation, restricting this cooling scheme to low-frequency mechanical oscillators. In contrast, collective spin states of atomic ensembles offer long coherence times and wide magnetic tuning of the spin precession frequency across the MHz range. Crucially, a versatile quantum toolbox exists that provides sophisticated techniques for ground-state cooling and quantum control \cite{hammerer2010,pezze2018}. This makes it possible to use the atomic spin as a coherent feedback controller, which can be employed to efficiently cool and control the mechanical oscillator \cite{vogell2015}, e.g., via a state-swap \cite{wallquist2010}.

Here, we demonstrate coherent feedback control of a nanomechanical membrane oscillator with the collective spin of an atomic ensemble and employ it to cool the membrane. For this, we exploit the coherent control offered by our recently demonstrated spin-membrane interface, where light mediates strong coupling between the two systems \cite{karg2020}. Using optical pumping on an internal atomic transition we can modify the spin damping rate and study the membrane cooling performance in different regimes. We show that coherent state swaps alternated with spin pumping pulses allow us to extract the noise from the mechanical system in an efficient way, providing the largest cooling rate and reaching the phonon steady-state faster than for continuous cooling. Finally, we study the effect of feedback delay onto the steady-state temperature of the membrane in the light-mediated coupling between the mechanical and spin systems. Our observations agree well with a theoretical model. 

\section{Setup}\label{sec:setup}

\begin{figure}
\includegraphics[width=\linewidth]{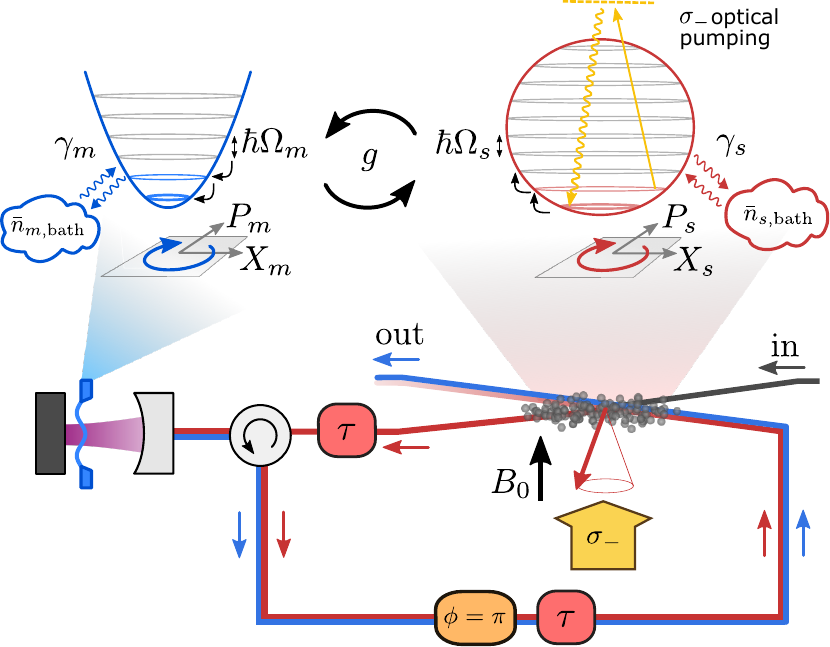}
\caption{\label{fig:epsart} Sketch of the light-mediated spin-membrane coupling. 
Light interacts first with the spin, then with the membrane, and then again with the spin. The propagation of the light leads to a feedback delay $\tau$. On the way back from the membrane to the spin, a $\pi$-phase is imprinted on the light, rendering the spin-membrane interaction effectively Hamiltonian for zero-delay $\tau=0$.  
The systems can be approximated by harmonic oscillators of frequencies $\Omega_m$ and $\Omega_s$ with damping rates $\gamma_m$ and $\gamma_s$ coupling them to a bath with $\bar n_{m,\mathrm{bath}}$ and $\bar n_{s,\mathrm{bath}}$ phonons, respectively. The oscillators are coupled at a rate $g$. The spin damping rate can be increased by applying a $\sigma_-$-polarized pumping laser.}
\label{fig: sketch of experiment}
\end{figure}

\begin{figure*}[t]
\includegraphics{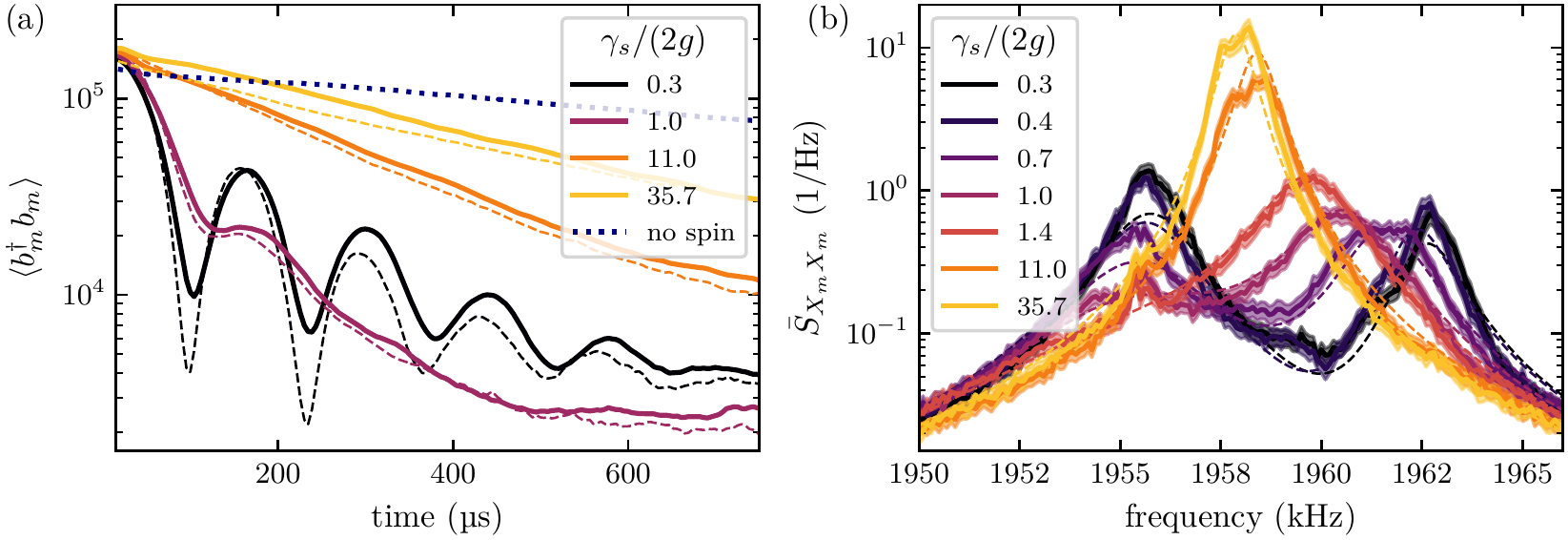}
\caption{(a): Time traces of the membrane occupation number after turning on the coupling to the atoms. The different traces show measurements with different spin damping rates $\gamma_s$. The dashed lines correspond to the simulation described in the text based on Eqs.~\eqref{eq.equations of motion 1} and \eqref{eq.equations of motion 2}. The dotted line shows the membrane dynamics without atoms but with the coupling beam turned on. {(b)}: Power spectral density of the membrane displacement. The dashed lines show a global fit to the data with the initial phonon occupation $\langle b^\dagger_i b^{}_i \rangle(t=0)$, $\Omega_m$, $\tau$, $g$, and the detector shot noise level as global fit parameters and $\Omega_s$ and $\gamma_s$ as individual fit parameters. All other parameters were taken from independent calibrations. In (a) and (b), solid lines correspond to the mean and shaded areas to the standard deviation of 355 measurements.}
\label{fig: continuous pump 2}
\end{figure*}

Our hybrid system consists of a mechanical oscillator and a collective atomic spin coupled by laser light over a distance of 1 meter in a loop geometry (Fig.~\ref{fig: sketch of experiment}). The mechanical oscillator is the (2, 2) square drum mode of a silicon-nitride membrane \cite{thompson2008}, which has a vibrational frequency $\Omega_m = 2\pi \times \SI{1.957}{\mega \Hz}$ and an intrinsic quality factor $Q=1.4\times10^6$. The membrane is placed in a single-sided optical cavity of linewidth $\kappa = 2\pi\times \SI{77}{\mega\Hz}$, which enhances the optomechanical coupling to external fields. The cavity is driven by an auxiliary laser beam (not shown in Fig.~\ref{fig: sketch of experiment}) that is red-detuned
from the cavity resonance, providing some initial cavity optomechanical cooling of the membrane to $2\times10^5$ phonons \cite{aspelmeyer2014}. The reflection of this beam is used to stabilize the cavity length and read out the membrane displacement via homodyne detection (detailed in Appendix \ref{sec:membrane Calibration}).

The collective spin is realised with an ensemble of $1.3\times 10^7$ cold ${}^{87}$Rb atoms confined in an optical dipole trap. Strong coupling of the atomic ensemble to the light is ensured by its large optical depth $d_0 \approx 300$. The atomic spins are optically pumped into the hyperfine ground state $|{F}=2,{m_F} = -2 \rangle$ with respect to a static magnetic field $ B_0 = \SI{2.8}{G}$ perpendicular to the propagation direction of the coupling laser. The Larmor frequency $\Omega_s \propto B_0$ is tuned into resonance with the membrane frequency $\Omega_m$. The spin precession is measured after the first interaction with the coupling laser by picking up a small fraction of the light (calibration shown in Appendix \ref{sec:spin Calibration}). The small-amplitude dynamics of the transverse spin components can be described by a harmonic oscillator of frequency $\Omega_s$ using the Holstein-Primakoff approximation \cite{hammerer2010}.

A coupling laser beam interacts first with the spin, then with the membrane, and once again with the spin, as sketched in Fig.~\ref{fig: sketch of experiment} and detailed in \cite{karg2020}. The coupling beam with \SI{1}{\milli\watt} optical power is slightly red-detuned with respect to the membrane cavity and $-2\pi\times\SI{40}{\giga\Hz}$ red-detuned from the ${}^{87}$Rb $\mathrm{D}_2$-line. It cools the membrane further to $\bar n_{m\mathrm{,bath}} = 2.0\times10^4 $ phonons, which broadens its linewidth to $\gamma_m = 2\pi \times \SI{262}{\Hz}$. In presence of the coupling beam, the spin linewidth is $\gamma_s = 2\pi \times \SI{2.2}{\kilo\Hz}$. In the first spin-light interaction, the $X_s$ quadrature of the atomic spin is imprinted onto the coupling beam via the Faraday interaction \cite{hammerer2010}, resulting in a modulation of the radiation-pressure force on the membrane. Likewise, the membrane displacement $X_m$ modulates the light reflected from the cavity \cite{aspelmeyer2014} which then creates a torque on the spin in the second interaction. On the way back from the membrane to the spin, the optical field carrying the spin and membrane signals is phase-shifted by $\pi$ such that the effective spin-membrane interaction is predominantly Hamiltonian and the backaction of the light on the spin is suppressed \cite{karg2019}. 
Tracing out the light field and neglecting the propagation delay for the moment, the resonant part of the effective spin-membrane interaction is described by a beam splitter Hamiltonian 
$H_\mathrm{BS} = \hbar g (b_s^\dagger b^{}_m + b_m^\dagger b^{}_s)$,
where $b_m$ ($b_s$) is the annihilation operator of a membrane (spin) excitation and $g$ is the effective spin-membrane coupling rate \cite{karg2020}. 

\section{Continuous Cooling}

Recently, we demonstrated strong coupling with this spin-membrane interface, i.e. $2g  > (\gamma_s + \gamma_m) \approx \gamma_s$ \cite{karg2020}. Strong coupling is manifested by the hybridization of the membrane and spin modes which leads to a normal mode splitting of $2g = 2\pi \times \SI{6.8}{\kilo\Hz}$ in the spectrum as shown in Fig.~\ref{fig: continuous pump 2}(b). In the time domain, strong coupling gives rise to state swaps between the spin and the membrane at the coupling rate $g$. In Fig.~\ref{fig: continuous pump 2}(a) we show the time evolution of the membrane occupation number after switching on the coupling beam. For $2g>\gamma_s$, the thermally excited membrane swaps its state with the spin, which is initially prepared close to its ground-state, in half a period $T_\mathrm{\pi}=\pi/g$ of the energy exchange oscillations. After another half period, the thermal state is swapped back onto the membrane but the phonon number is reduced due to the damping that occurred in the spin system, whose linewidth is larger than that of the membrane. The oscillations dephase after approximately $ \SI{1}{\milli\second}$ and a steady state with a membrane occupation of $\bar n_{m,\mathrm{ss}} \approx 2.3\times10^3$ phonons is reached, corresponding to a temperature decrease by two orders of magnitude compared to the initial state. In this process the membrane is predominantly cooled via its coupling to the cold and damped spin, reaching a temperature one order of magnitude lower than in the presence of the optomechanical cooling beams alone. 

We now study the effect of increasing the spin damping rate $\gamma_s$ on the coupled dynamics. To increase $\gamma_s$ we apply a $\sigma_-$-polarized pump laser along the polarization axis of the spin (calibration in Appendix \ref{sec:gamma_s Calibration}). As can be seen in Fig.~\ref{fig: continuous pump 2}(a), increasing $\gamma_s$ first enhances the membrane cooling, until the overdamped regime $\gamma_s \gg 2g$ is reached where the membrane couples incoherently to a quasi-continuum of cold spin fluctuations. The membrane decay is then governed by Fermi's golden rule, with the occupation number decreasing at the sympathetic cooling rate $\gamma_\mathrm{sym} \approx 4g^{2}/\gamma_s$, i.e.\ the cooling becomes less effective as $\gamma_s$ is increased further. In this weak-coupling regime, the modes decouple and the membrane spectrum shows a single Lorentzian peak, broadened by the interaction with the spin, see Fig.~\ref{fig: continuous pump 2}(b).

\section{Stroboscopic Cooling}

\begin{figure}[t]
\includegraphics{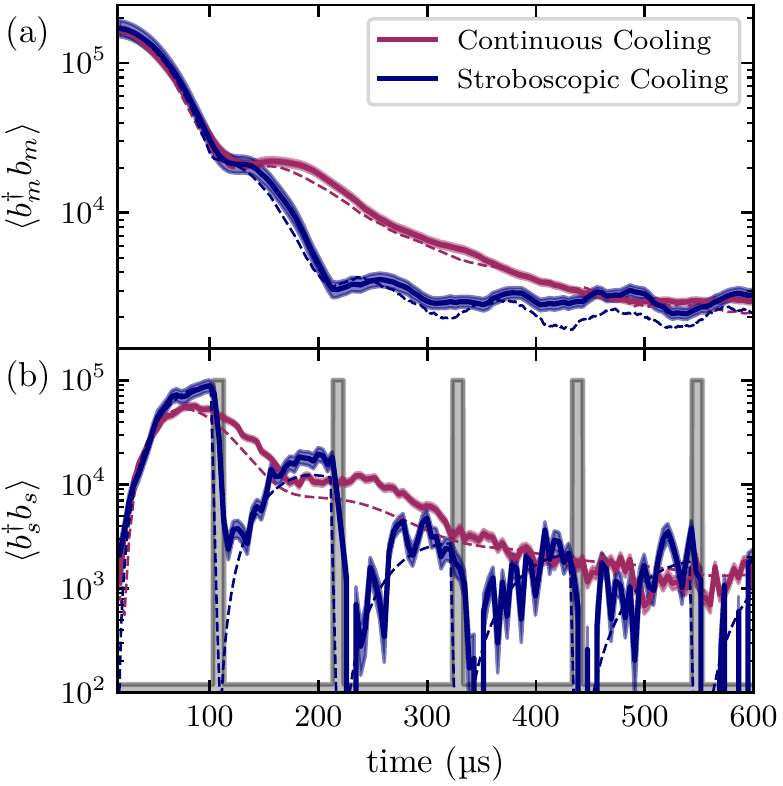}
\caption{(a) Membrane and (b) spin occupation numbers for continuous cooling at $\gamma_s = 2g$ and stroboscopic cooling at $\gamma_s = 0.6g$. The gray shaded areas indicate the spin pumping pulses (where $\gamma_s \approx 60g$). Solid lines and shaded areas correspond to the mean and standard deviation of 70 measurements and dashed lines correspond to a simulation.}
\label{fig: pulsed pump}
\end{figure}

Previous experiments, which coupled a membrane to the motion of cold atoms \cite{jockel2015,christoph2018}, lacked both strong coupling and coherent control over the atoms. In contrast, our strongly coupled spin-membrane system allows us to implement more elaborate coherent control schemes. In particular, we can combine strong coupling and strong spin damping in a stroboscopic fashion in order to cool the membrane much faster than in the continuous cooling case discussed above. In Fig.~\ref{fig: pulsed pump} we show a comparison between stroboscopic and continuous cooling, where time traces for (a) the membrane occupation number and (b) the spin occupation number are shown. In the stroboscopic sequence we perform a coherent $\pi$-pulse ($T_\mathrm{pulse} = \SI{100}{\micro\second}$, $\gamma_s = 0.6g$) to swap membrane and spin states. Afterwards, we apply an optical pumping pulse of duration $T_\mathrm{pump} = \SI{10}{\micro\second}$ which increases the spin damping rate to $\gamma_s \approx 60g$ and depletes the spin occupation on a timescale much shorter than the state swap (gray pulses in Fig.~\ref{fig: pulsed pump}(b)). During the pumping pulse the coupling is kept on. Since the spin is reinitialised close to the ground state, the next coherent state swap does not transfer thermal energy back to the membrane but only cools it further. It takes two to three such iterations of a coherent $\pi$-pulse followed by a spin pumping pulse to reach the steady state (see Fig.~\ref{fig: pulsed pump}). Using this simple sequence, we can reach the membrane steady state temperature of \SI{216}{\milli\kelvin} ($\bar{n}_{m,\mathrm{ss}} = 2.3\times 10^3$  phonons) in around \SI{200}{\micro\second}, approximately a factor of two faster than for continuous cooling. This exemplarily shows the advantage of a coherent feedback controller, which enables faster cooling than if the membrane is coupled with a similar rate to an incoherent, overdamped system.

\begin{figure}[t]
\includegraphics{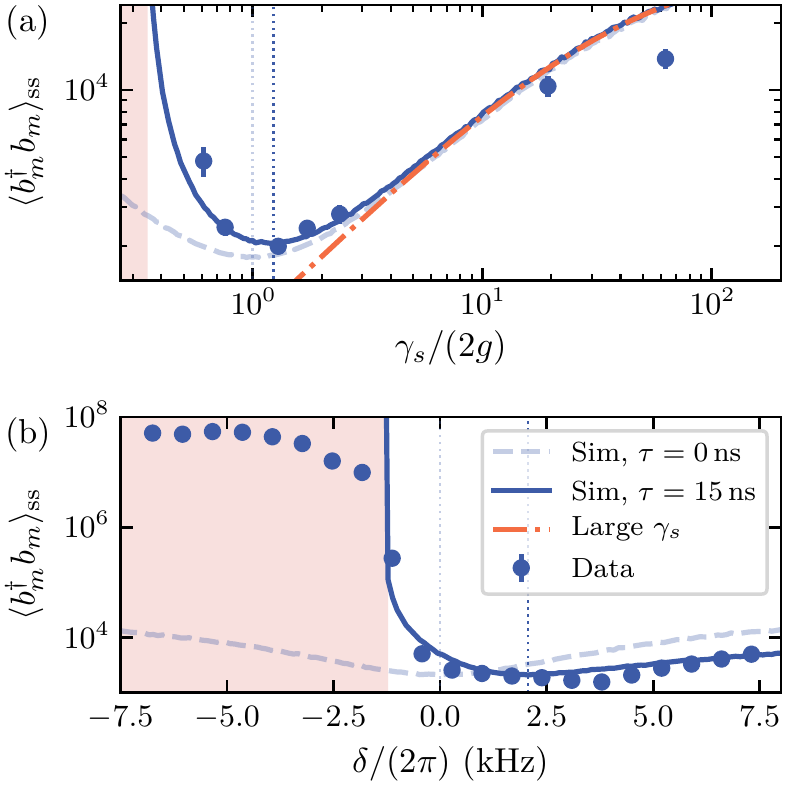}
\caption{Steady state occupation of the membrane as a function of (a) spin damping rate $\gamma_{s}$ (at resonance, $\delta = 0$) and (b)  spin-membrane detuning  $\delta = \Omega_s - \Omega_m$ at $\gamma_s = 0.6g$. The solid (dashed) blue line shows the result of the simulation with (without) delay. In (a), the red dashed-dotted line indicates the steady-state number given by the rate in Eq.~\eqref{eq:delay_gamma_s} with $\tau = \SI{15}{\nano\second}$. The red shaded area shows the region for which the dynamics is found to be unstable using the Routh-Hurwitz criterion. For this measurement,  $\bar n_{m,\mathrm{bath}} \approx 4.0 \times 10^4$ phonons and $\gamma_m = 2\pi\times\SI{94}{Hz}$ (independently calibrated without atoms). The data points with error bars correspond to the mean and the standard deviation of steady state occupations of 20 (3) experimental realisations in (a) ((b)).}
\label{fig:steady_state_vs_gamma_s}
\end{figure}

\section{Theoretical Model}

Further insight into the dynamics is gained by solving the equations of motion for the coupled spin-membrane system \cite{karg2020},
\begin{align}
    &\ddot{X}_{m} + \gamma_{m} \dot{X}_{m} + \Omega_{m}^2 X_{m} = - 2g \Omega_{m} X_{s}(t-\tau) + \mathcal{F}_m \label{eq.equations of motion 1},\\
    &\ddot{X}_{s} + \gamma_{s} \dot{X}_{s} + \Omega_{s}^2 X_{s} = - 2g \Omega_{s}  X_{m}(t-\tau) + \mathcal{F}_s
    \label{eq.equations of motion 2},
\end{align}
where terms on the left-hand-side describe the internal dynamics of the damped oscillators and the first term on the right-hand-side describes the state swap dynamics including a propagation delay $\tau$ between the spin and the membrane. We included the generalized Langevin forces $\mathcal{F}_m$ and $\mathcal{F}_s$ that capture stochastic force terms due to quantum fluctuations, thermal and measurement backaction noise (detailed in Appendix \ref{sec:Simulation Model}).

We used the following procedures to simulate our experimental results: for the continuous cooling measurements, we first fitted the spectra for different $\gamma_s$ in Fig.~\ref{fig: continuous pump 2}(b) globally using a coupled-mode model (fit function given in Appendix \ref{sec:PSD coupled-mode}). From this fit, the extracted $\tau$ and $\Omega_m$ were used as the input parameters for the simulation. We adapted the technique described in \cite{norrelykke2011} to numerically solve the equations of motion \eqref{eq.equations of motion 1} and \eqref{eq.equations of motion 2} and compare the solution to our data (more details are given in Appendix \ref{sec:Simulation Model}). To generate each time trace in Fig.~\ref{fig: continuous pump 2}(a) (dashed lines) we fitted the numerical solution to our data with only $\gamma_s$ and $\Omega_s$ as free parameters. The fit results show a systematic shift of $\Omega_{s}$ with increasing spin pumping power, likely due to the light shift induced by the circularly polarised pumping laser  (Fig.~\ref{fig:spin_summary}), and $\gamma_{s}$ was observed to be larger than in the independent calibration of Appendix \ref{sec:gamma_s Calibration}.
 
For the stroboscopic cooling measurements, we took the fit parameters from the continuous cooling measurement and ran the simulation with a time dependent spin damping rate which was taken to be $\gamma_s = 0.6g$ during the state swaps and $\gamma_s = 60g$ during the pumping pulses. The fit is shown for membrane and spin in Fig.~\ref{fig: pulsed pump} as a dashed line. The good agreement between fit and data shows that our model includes all the relevant factors which govern the coupled dynamics.

\section{Delayed Feedback}

Our hybrid spin-membrane system constitutes a coherent feedback network \cite{bennett2014}, in which delayed feedback can give rise to instabilities \cite{reddy1998,reddy2000,vochezer2018}. In our experiment, such instabilities show up as a spontaneous coupled oscillation of spin and membrane, which we observe for certain values of the spin-membrane detuning $\delta = \Omega_s - \Omega_m$. Even at resonance, we have to include the feedback delay to predict the experimentally measured steady state occupation of the membrane accurately. In Fig.~\ref{fig:steady_state_vs_gamma_s} we plot the measured and simulated occupation numbers of the membrane in steady state as a function of $\gamma_s$ [Fig.~\ref{fig:steady_state_vs_gamma_s}(a)] and $\delta$ [Fig.~\ref{fig:steady_state_vs_gamma_s}(b)]. At resonance and for $\Omega_m \tau \ll 1$ (as in our system), the effect of the feedback delay is most apparent in the limit of small $\gamma_s$. The model without delay (light-blue dashed line) predicts a significantly smaller occupation number compared to both what we observe in experiments and what is predicted by our model including the feedback delay (blue solid line). In the large $\gamma_s$ limit, the sympathetic cooling rate is modified to
\begin{equation}
\gamma_\mathrm{sym} \approx \frac{4g^{2}}{4\delta^2 + \gamma_s^2}\left[\gamma_s\cos(2\Omega_m \tau) + 2\delta\sin(2\Omega_m \tau)\right]
\label{eq:delay_gamma_s}
\end{equation}
(see Appendix \ref{sec:sympathetic cooling} for derivation). In this limit, the steady state occupation is given asymptotically by $ \langle b^\dagger_m b^{}_m\rangle_\mathrm{ss} = \bar n_{m,\mathrm{bath}} \gamma_m / (\gamma_m + \gamma_\mathrm{sym})$, shown as the red dashed-dotted line in Fig.~\ref{fig:steady_state_vs_gamma_s}(a). The theory of coupled oscillators without delay predicts optimal sympathetic cooling at the critical damping of $\gamma_s = 2g$ (faded vertical dotted line in Fig.~\ref{fig:steady_state_vs_gamma_s}). Including the feedback delay in the model, the minimal occupation number shifts to larger $\gamma_s$ (dark vertical dotted line), because the self-oscillations have to be compensated by a higher spin damping rate. The experimental data confirms this theoretical prediction.

Furthermore, we find that the presence of delay lifts the symmetry in $\delta$, as inferred theoretically from Eq.~(\ref{eq:delay_gamma_s}) for large $\gamma_s$ and shown both experimentally and theoretically in Fig.~\ref{fig:steady_state_vs_gamma_s}(b) for small $\gamma_s=0.6g$. We see that the minimal steady state occupation of the membrane is obtained for positive detuning $\delta$, i.e. $\Omega_s>\Omega_m$, which is true in general for a feedback system with a delay of $\tau < \pi/(2\Omega_m)$. For large enough negative $\delta$, we observe that the coupling drives the system into limit cycle oscillations, see Fig.~\ref{fig:steady_state_vs_gamma_s}(b). With our model we can attribute these self-oscillations to the feedback delay. In this self-driven regime, the resulting membrane occupation of $6.8\times 10^7$ exceeds the spin length by around a factor of three. The emergence of such instabilities can be characterised using the Routh-Hurwitz stability criterion \cite{hofer2008}, which indicates whether the real part of one of the normal modes of the system reverses its sign (shown in Appendix \ref{sec:Hurwitz}). In Fig.~\ref{fig:steady_state_vs_gamma_s} we indicate such unstable regions for our coupled system by a shaded area. Our calculations show that the precise value of $\delta$ at which the driving due to the loop delay exceeds the damping of the coupled system depends on  $\gamma_s$. Even at resonance [Fig.~\ref{fig:steady_state_vs_gamma_s}(a)] self-oscillations are predicted for small enough $\gamma_s$. 

The propagation delay is an interesting tuning knob for coherent feedback experiments, which gives access to Hamiltonian and dissipative dynamics: We can induce self-oscillations of the system, tune the dependence of the steady state on system parameters such as damping rate and detuning, or even render the delay negligible by tuning $2\Omega_m \tau$ to a multiple of $2 \pi$.

\begin{figure}[tb]
\includegraphics{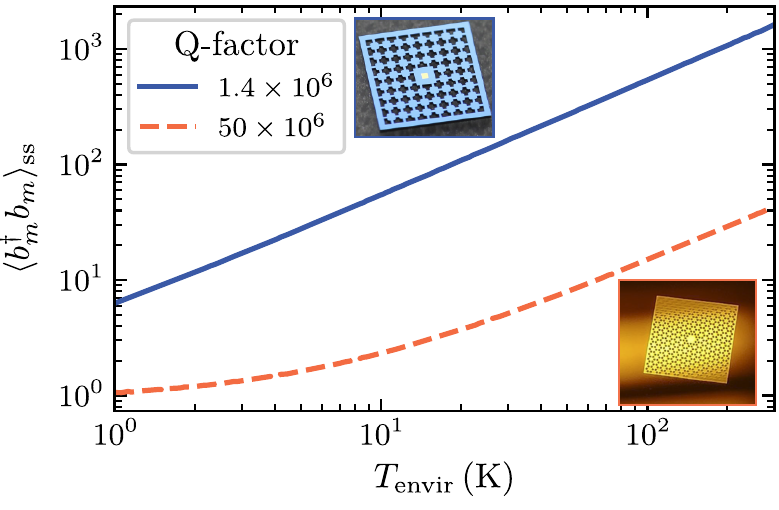}
\caption{Simulated steady state occupation of membranes for varying cryostat temperature and different mechanical $Q$ factors. Here,  $\gamma_s = 2g$,  $\delta = 0$ and $\tau = \SI{15}{\nano\second}$. The insets show the current membrane with phononic shield used in these experiments and a soft-clamped membrane for which $Q \approx 5\times 10^7$.}
\label{fig:steady_state}
\end{figure}

\section{Discussion}
In our experiment, the cooling rate of the membrane due to its coupling to the spin exceeds the cavity-optomechanical cooling rate by more than one order of magnitude. The lowest achievable phonon occupation of the membrane is thus given by the competition of cooling the membrane with the spin and heating due to its coupling to the room-temperature environment. In Fig.~\ref{fig:steady_state} we show the expected membrane steady state occupation for varying environment temperature and two different membrane designs. In this calculation we include the cavity-optomechanical cooling of the membrane (which has a negligible effect), the light-mediated coupling to the spin including backaction of the light, as well as thermal and quantum mechanical ground state fluctuations of both systems. The higher quality factors  $Q > 5 \times 10^7$ of soft-clamped membranes \cite{tsaturyan2017, reetz2019} would reduce the thermal decoherence rate by a factor 25 and allow us to prepare the mechanical oscillator close to its ground state in a  \SI{4}{\kelvin} environment. These technical improvements would realize a mechanical oscillator whose phonon occupation is limited by quantum backaction instead of thermal noise. While in the current coupling scheme the double pass eliminates backaction on the atomic spin, a large membrane quantum cooperativity $ C_{m} > 1 $ would favor a double pass scheme with coherent cancellation of quantum backaction on the membrane. This would lead to a higher quantum cooperativity for the spin-membrane coupling \cite{karg2019}. Further, the feedback control of the membrane could be improved by increasing the quantum cooperativity of the spin system. This involves gaining a better understanding of the spin decoherence sources and achieving a larger spin-light coupling rate.

In this work we implemented a relatively simple coherent feedback sequence based on coherent state swaps of pulse area $\pi$ interleaved with short spin pumping pulses. In the future, it would be interesting to explore more elaborate feedback sequences to optimize the cooling in a specific situation. For example, the duty cycle of the stroboscopic cooling sequence could be changed over time to cool a mechanical oscillator with a high initial occupation that exceeds the spin length. Initially, short coupling pulses of pulse area $\ll \pi$ could remove excitations without saturating the spin, and once the phonon number is sufficiently reduced, the pulse area could be increased to minimize the final temperature.

Our coherent feedback cooling scheme is a rather general technique that can be applied to any physical system with a strong light-matter interface.  This includes cavity optomechanical systems or mechanical oscillators without an optical cavity. Moreover, similar cooling schemes could be implemented in the microwave domain with electromechanical oscillators \cite{clerk2020} coupled to solid-state spin systems. The macroscopic distance between the feedback controller and the target system enables modular control schemes in analogy to classical feedback in electrical engineering. This  opens up the new possibility to use coherent feedback control in quantum networks.

The coherent control and bidirectional Hamiltonian coupling employed in this work pave the way towards more elaborate quantum protocols such as the generation of non-classical mechanical states via state swaps \cite{wallquist2010} as well as further studies of coherent feedback in the quantum regime \cite{lloyd2000,wiseman2009,hamerly2012, zhang2017}.

\begin{acknowledgements}
We thank Christoph Bruder for a careful reading of the manuscript. This work was supported by the project “Modular mechanical-atomic quantum systems” (MODULAR) of the European Research Council (ERC) and by the Swiss Nanoscience Institute (SNI). MBA acknowledges funding from the European
Union’s Horizon 2020 research and innovation programme under the Marie Sk\l odowska-Curie
grant agreement N°101023088.
\end{acknowledgements}

\newpage
\appendix
\section{Calibrations}

\subsection{Calibration of the spin signal}
\label{sec:spin Calibration}
The spin occupation was calibrated by using the off-resonant Faraday interaction \cite{hammerer2010} between the atoms and the light. In contrast to the looped coupling scheme in which light interacts twice with the spin, the calibration is performed by measuring the light directly after the first interaction with the spin. Hence, the light does not interact with the membrane nor with the atoms a second time. 

\begin{figure}[h]
\includegraphics{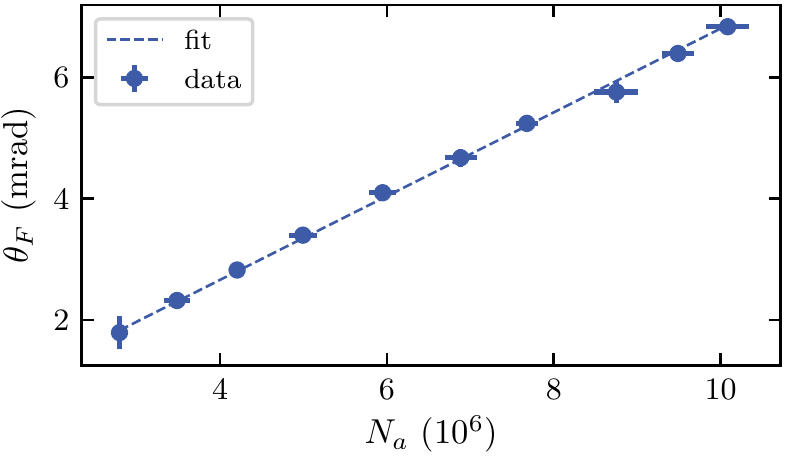}
\caption{DC Faraday rotation signal of the atomic ensemble spin-polarized along the optical propagation axis at a probe laser detuning of $-2\pi\times\SI{40}{\giga\hertz}$. In the experiment, first the Faraday rotation angle $\theta_F$ was measured with a weak, far-detuned probe pulse and then the number of atoms $N_a$ was determined by absorption imaging. Each data point with error bar corresponds to the mean and the standard deviation of five experimental runs at the same MOT loading time.}
\label{fig:DC_faraday}
\end{figure}

For a single spin-light interaction, the Hamiltonian of the Faraday interaction is given by \cite{geremia2006} 
\begin{equation}
H_\mathrm{int} = \hbar \alpha_1 S_z F_z
\end{equation}
where $\alpha_1$ is the vector polarizability of the atoms, $S_z$ is the circularly polarized component of the Stokes vector of the light, and $F_z$ is the collective spin component along the propagation direction of the probe laser. The input-output relation for the $S_y$ Stokes vector component of the probe light yields
\begin{equation}
    S_y^{(\mathrm{out})} = S_y^{(\mathrm{in})} + \alpha_1 S_x^{(\mathrm{in})} F_z .
\end{equation}
For the experiments in this paper, the probe laser was linearly polarized with an angle of $55^\circ$ with respect to the magnetic field in order to minimise frequency shifts due to the tensor interaction with the light \cite{geremia2006}, which otherwise would give rise to inhomogeneous broadening of the spin. In the following we define $S_x$ as the difference between the flux of light linearly polarized at $55^\circ$  and the flux in the orthogonal linear polarization, such that we can approximate the $S_x$ operator as a classical quantity $S_x^{(\mathrm{in})} \approx S_x^{(\mathrm{out})} \approx \langle S_x \rangle$ and define the Faraday angle $\theta_F = S_y^{(\mathrm{out})}/(2\langle S_x \rangle) = \alpha_1 F_z/2$. For the calibration measurement, we slowly rotate the spin to align it to the propagation direction of light. Thus, the collective atomic spin points along the $z$-axis, and the $F_z$ component of the spin can be approximated by $F_z \approx \langle F_z \rangle = {F} N_a = 2 N_a$ where the number of atoms in the dipole trap $N_a$ was measured independently by absorption imaging. The $S_y^{(\mathrm{out})}$ Stokes vector component of the out-going field was determined by a polarization homodyne measurement. Measuring $\langle S_x \rangle$ independently allows us to determine the Faraday angle (shown in Fig.~\ref{fig:DC_faraday} for different atom numbers). By knowing the number of atoms from absorption imaging and the Faraday angle from polarization homodyne measurement, the ensemble-averaged vector polarizability $\alpha_1 = \theta_F/N_a$ can be calculated.\\

With this calibrated value for the vector polarizability $\alpha_1$, a measurement of the Faraday angle $\theta_F$  yields directly the $F_z$ component of an arbitrary spin state. If the collective spin is aligned along a magnetic field perpendicular to the propagation direction of the light field, the $F_z$ component oscillates at the Larmor frequency. In this case of an oscillating signal, the polarization homodyne measurement of the out-going field is demodulated by a lock-in amplifier which returns the  root mean square (rms) amplitude $V_{S_{y},\mathrm{rms}}^\mathrm{\SI{50}{\ohm}} \propto \bar S_{y}$ (where $\bar S_{y}$ is the slowly varying amplitude of $S_{y}$). In order to determine the Faraday angle from this rms amplitude and the DC measurement of $V_{S_{x}} \propto \langle S_x \rangle$ using an oscilloscope, one has to first multiply the rms amplitude by a factor $\sqrt{2}$ to get a peak amplitude voltage and further by a factor of $2$ to compensate for the impedance mismatch between the \SI{50}{\ohm} input impedance of the lock-in amplifier and the high input impedance of the oscilloscope. Including these factors, we get the slowly varying amplitude of the Faraday angle $\bar \theta_F = \sqrt{2}V_{S_{y},\mathrm{rms}}^\mathrm{\SI{50}{\ohm}}/V_{S_{x}}$. By normalising the spin signal by the square-root of the total spin length we obtain the slowly varying amplitude of the $X_s$-quadrature of the spin
\begin{equation}
    \bar X_s = \frac{\bar F_z}{\sqrt{\langle F_x \rangle}} = \frac{2\bar\theta_F}{\alpha_1\sqrt{2N_a}} = \frac{2V_{S_{y},\mathrm{rms}}^\mathrm{\SI{50}{\ohm}}}{\alpha_1 V_{S_{x}}\sqrt{N_a}}.
\end{equation}
From this result, we can use the equipartition theorem to calculate the number of excitations of the spin oscillator:
\begin{equation}
    \bar n_s + \frac{1}{2} = \frac{\langle X_s(t)^2 + P_s(t)^2\rangle}{2} = \bar X_s^2 = \frac{2\bar \theta_F^2}{\alpha_1^2 N_a}.
\end{equation}
where $X_s(t)$ and $P_s(t)$ are the fast rotating quadratures of the spin oscillator. In the looped experiment, only a small fraction of the light was measured in between the first interaction of the spin and the interaction of the light with the membrane. This in-loop measurement was calibrated using a coherent spin excitation and comparing it to the polarization homodyne measurement presented in this section.

\subsection{Calibration of the spin damping rate}
\label{sec:gamma_s Calibration}
\begin{figure*}[t]
\centering
\includegraphics{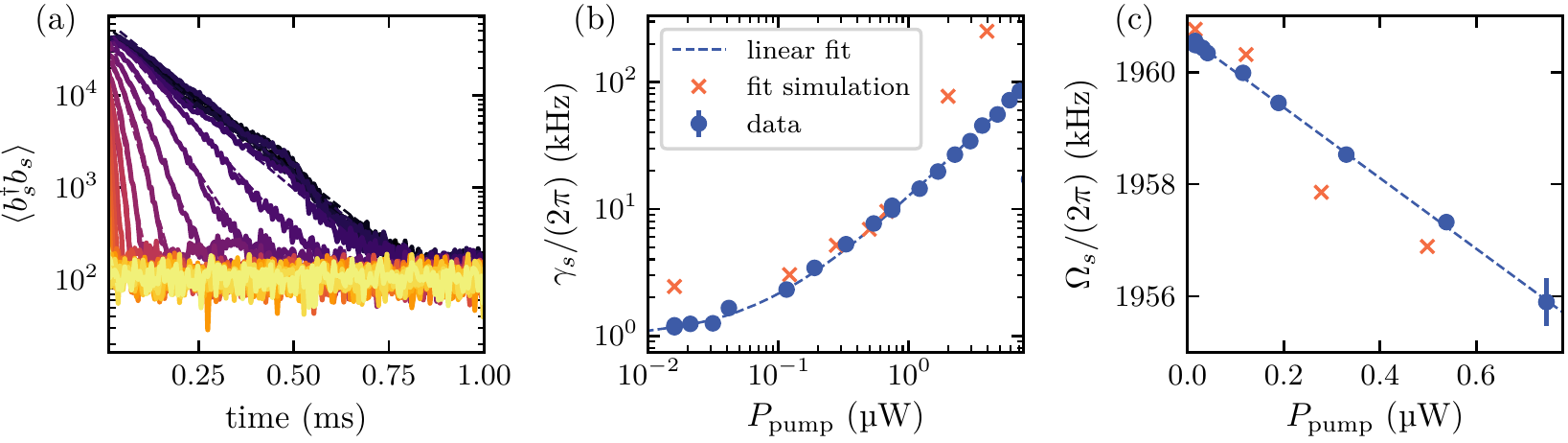}
\caption{Measurement of the spin in the absence of coupling to the membrane after it is excited by a weak RF-pulse: (a) Time trace of double pass measurement of the spin with different pumping powers (range from $0$ to \SI{10}{\micro\watt}). The dashed lines show fits with an exponential decay.  The spin linewidth (b) and spin frequency (c) are plotted as a function of the pump power. The dashed lines in (b) and (c) show a linear fit to the spin linewidth and resonance frequency. The crosses show the fit parameters extracted from Fig.~\ref{fig: continuous pump 2}(a) which were used as input for the simulations. The data shown in (a) is an average over seven experimental realisations and was used to fit the exponential decay [for (b)] and the Lorentzian peak [for (c)]. The error bars in (b) and (c) show the fit-error of the corresponding quantity.}
\label{fig:spin_summary}
\end{figure*}

One of the main parameters in the experiments is the spin damping rate $\gamma_s$. In order to measure the spin damping rate in the presence of all lasers but without coupling to the membrane, we detuned the coupling laser from the cavity resonance ($|\Delta|\gg\kappa$). The laser thus is reflected from the incoupling mirror of the cavity and only the spin is probed. For the calibration measurements, the spin was coherently excited by a weak RF-pulse. The spin signal was measured by detecting the remaining signal on the light after the second pass. It is normalised to occupation numbers [shown in Fig.~\ref{fig:spin_summary}(a)]. The damping rate $\gamma_{s}$ is extracted from the exponential fit to the temporal dynamics [Fig.~\ref{fig:spin_summary}(b)] and the frequency $\Omega_s$ is extracted from a Lorentzian fit to the spectrum [Fig.~\ref{fig:spin_summary}(c)]. For optical pumping power larger than $P_\mathrm{pump} > \SI{0.7}{\micro\watt}$, the spectra were too broad to provide reasonable fit results [and are therefore not shown in Fig.~\ref{fig:spin_summary}(c)]. In Fig.~\ref{fig:spin_summary}(b) and (c), fit parameters for the coupled dynamics are shown.

\subsection{Calibration of the membrane signal}
\label{sec:membrane Calibration}
The vibrations of the membrane are detected via their effect on the phase of the beam reflected from the cavity \citep{aspelmeyer2014}. In particular, the membrane vibrations modulate the cavity resonance frequency $\omega_c$. For small membrane displacements, we can write $\omega_c(x_m) \approx \omega_c + Gx_m$, where $G = -\mathrm{d}\omega_c/\mathrm{d} x_m$ is the cavity frequency shift per membrane displacement. For a single-sided cavity, the phase $\phi_c$ of the beam reflected from the cavity with respect to the incoming beam is related to the cavity detuning $\Delta = \omega_L-\omega_c$ by \citep{aspelmeyer2014}
	\begin{equation}
	\phi_c = \arctan\left[\frac{\kappa\Delta}{(\kappa/2)^2+\Delta^2}\right],
	\end{equation}	 
where $\kappa$ is the cavity linewidth. A change in the cavity frequency $\delta\omega_c = Gx_m$ thus leads to a change in the phase of the reflected beam by an amount $\delta\phi_c = -(\mathrm{d}\phi_c/\mathrm{d}\Delta)\delta\omega_c\approx -{4}G x_m/{\kappa}$, where the approximation holds for  small detunings $|\Delta|\ll\kappa$. We can write the previous expression in terms of the vacuum optomechanical coupling strength $g_0 = Gx_\mathrm{ZPF}$ as
	\begin{equation}
	\delta\phi_c = -\frac{4g_0}{\kappa}\frac{x_m}{x_\mathrm{ZPF}},
	\end{equation}
where we have introduced the zero-point fluctuation amplitude of the membrane $x_\mathrm{ZPF} = (\hbar/2m_\mathrm{eff}\Omega_m)^{1/2}$, with $m_\mathrm{eff}$ the effective mass of the vibration mode. These phase variations $\delta\phi_c$ can now be read interferometrically by means of balanced homodyne detection. For this, the beam reflected from the cavity is combined with a strong local oscillator in a 50:50 beam splitter. The output beams are subsequently photodetected and the output signals subtracted. The recorded balanced voltage can be written as 
	\begin{equation}
	V = V_0\cos(\Delta\phi),
	\end{equation}
with $V_0$ the modulation amplitude, proportional to the square-root of the power of the beam reflected from the cavity and of the local oscillator beam and with $\Delta\phi = \phi_c - \phi_\mathrm{LO} $ where $\phi_\mathrm{LO}$ is the phase of the local oscillator. 

The modulation amplitude $V_0$ is inferred by modulating $\phi_\mathrm{LO}$, thanks to a movable mirror in the local oscillator path which allows to generate path differences of a few wavelengths. $V_0$ can be extracted from the contrast of the observed interference fringes. In order to detect the phase fluctuations $\delta\phi_c$ of $\phi_c$ induced by the membrane motion, we lock the relative phase $\Delta\phi$ to $\pi/2$, i.e., the point where the slope of the fringes is maximal. For small shifts $\delta\phi_c\ll\pi/2$, the recorded voltage variation $\delta V(t)$ is directly proportional to $\delta\phi_c(t)$, and thus to $x_m(t)$. In practice, $\delta V(t)$ is effectively increased by a factor $1/\eta_c$ due to imperfect cavity coupling, such that
\begin{equation}
x_m(t) = \frac{\delta V(t)}{\eta_c V_0}\frac{x_\mathrm{ZPF}\kappa}{4g_0}.
\end{equation}	
In order to determine membrane phonon occupation $\bar n_m(t)$ we first define the dimensionless membrane quadrature operators $X_m = x_m/(\sqrt{2}x_\mathrm{ZPF})$ and $P_m = \sqrt{2}x_\mathrm{ZPF} p_m/\hbar$, defined so that $[X_m,P_m] = \mathrm{i}$. We can now write 
\begin{eqnarray}
\langle H(t)\rangle &&= \hbar\Omega_m\frac{\langle X_m(t)^2+P_m(t)^2\rangle}{2} \nonumber\\ &&=\hbar\Omega_m\left(\bar{n}_m(t)+\frac{1}{2}\right).
\end{eqnarray}
By means of the equipartition theorem, we can write $\langle X_m(t)^2\rangle_t=\langle P_m(t)^2\rangle_t$ and thus relate the measured voltage variations to the membrane phonon occupation number. In practice, we do not measure the voltage $\delta V(t)$ but its rms value $\delta V(t)^{50\,\Omega}_\mathrm{rms}$, which we further need to multiply by a factor of 2 due to impedance mismatch of our measuring instrument. To convert the measured rms value to amplitude variations, we thus need an overall $2\sqrt{2}$ factor.  This finally yields
\begin{eqnarray}
\bar{n}_m(t) + \frac{1}{2} &&= \langle X_m(t)^2\rangle_t\nonumber\\ &&=\left(\frac{\delta V(t)^{50\,\Omega}_\mathrm{rms}}{\eta_c V_0}\right)^2\left(\frac{\kappa}{2g_0}\right)^2.
\end{eqnarray}
The values of $\kappa = 2\pi\times\SI{77}{\mega\Hz}$ and $g_0 = 2\pi\times\SI{224}{\Hz}$, have been independently calibrated from the width of the Pound-Drever-Hall signal and by measuring the optomechanical response to an optical amplitude modulation tone, respectively.	

\section{Theoretical model for the membrane-spin coupling}
We modelled our membrane-spin coupling by two coupled harmonic oscillators as shown in Eqs.~\eqref{eq.equations of motion 1} and \eqref{eq.equations of motion 2}. In the following we show how we characterised, simulated, and approximated the system starting from these equations. In section \ref{sec:Simulation Model} the stochasic simulation of the system is presented. In section \ref{sec:PSD coupled-mode} we show the derivation of the fit function for the spectra. From the spectrum, we calculate the sympathetic cooling rate and the resonance frequency shift in the weak coupling limit in section \ref{sec:sympathetic cooling}. Finally, we show the Routh-Hurwitz stability analysis of the coupled dynamics with delay in section \ref{sec:Hurwitz}.

\subsection{\label{sec:Simulation Model} Simulation of the spin-membrane dynamics}
In this section, we provide some details on the simulation method we used to solve the stochastic equations of motion Eqs.~\eqref{eq.equations of motion 1} and \eqref{eq.equations of motion 2} for the spin-membrane system. This simulation follows closely the algorithm presented in \cite{norrelykke2011}. For the simulation, we rewrite the equations of motion as four coupled first-order differential equations for $\tilde{X}_j$ and $\tilde{P}_j$, with $j\in(m,s)$ in a frame rotating at the membrane frequency $\Omega_{m}$ (operators in the rotating frame are denoted with a tilde) and apply the rotating wave approximation (RWA). In the limit where the propagation delay is small compared to other timescales involved in the coupled dynamics (i.e. $ \tau \ll \gamma_j^{-1}, g^{-1}, \delta^{-1}$), the change of the oscillator quadratures during the time $\tau$ can be neglected in the rotating frame i.e. $\tilde{X}_j (t) \approx \tilde{X}_j (t-\tau)$ and $\tilde{P}_j (t) \approx \tilde{P}_j (t-\tau)$. The equations of motion then read

\begin{equation}
\frac{\mathrm{d}}{\mathrm{d} t}\begin{pmatrix}
\tilde{X}_{m}(t) \\
\tilde{P}_m(t) \\
\tilde{X}_s(t) \\
\tilde{P}_s(t) \\
\end{pmatrix}=-\mathbf{M}\begin{pmatrix}
\tilde{X}_m(t) \\
\tilde{P}_m(t) \\
\tilde{X}_s(t) \\
\tilde{P}_s(t) \\
\end{pmatrix}+\begin{pmatrix}
-\sin{(\Omega_{m}t)}\mathcal{F}_{m}(t) \\
\cos{(\Omega_{m}t)}\mathcal{F}_{m}(t) \\
-\sin{(\Omega_{m}t)}\mathcal{F}_{s}(t) \\
\cos{(\Omega_{m}t)}\mathcal{F}_{s}(t)
\end{pmatrix},
\label{1:EOM}
\end{equation}
where we have split the dynamics into  the $4 \times 4$ dynamical matrix 
\begin{widetext}
\begin{equation}
\mathbf{M}=\begin{pmatrix}
{\gamma_m}/{2} & 0 & -g \sin \left(\Omega_{m} \tau\right) & -g \cos \left(\Omega_{m} \tau\right) \\
0 & {\gamma_m}/{2}  & g \cos \left(\Omega_{m} \tau\right) & -g \sin \left(\Omega_{m} \tau\right) \\
- g \sin \left(\Omega_{m} \tau\right) & - g \cos \left(\Omega_{m} \tau\right) & {\gamma_s}/{2}   & -\delta\\
g \cos \left(\Omega_{m} \tau\right) & - g \sin \left(\Omega_{m} \tau\right) & \delta & {\gamma_s}/{2} 
\end{pmatrix},
\end{equation}
\end{widetext}
and a stochastic part, given by the generalized noise forces $\mathcal{F}_{j}(t) = \sqrt{2\gamma_{j}} F^\mathrm{(tot)}_{j}(t)$. The total force noise $\mathcal{F}_{j}^\mathrm{(tot)}(t)$ includes the thermal noise $F_{j}^{\mathrm{(th)}}(t)$ and the backaction noise $F^{\mathrm{(ba)}}_{j}(t)$ which itself depends on the optical vacuum noise $F^\mathrm{(in)}_{j}(t)$. Thus, it is given by
\begin{eqnarray}
F_{j}^\mathrm{(tot)}(t)&&=F_{j}^{\mathrm{(th)}}(t) +  F_{j}^{\mathrm{(ba)}}(t)\nonumber\\
&&=F_{j}^{\mathrm{(th)}}(t) + \sqrt{\frac{2 \Gamma_{j}}{\gamma_{j}}} F_{j}^\mathrm{(in)}(t),
\label{eq:generalized noise force}
\end{eqnarray}
where $\Gamma_{j}$ is the measurement rate of the individual system. The noise terms $F^{(\nu)}_{j}(t)$, $\nu\in\mathrm{(th, in)}$ can be expressed explicitly in terms of the product of a noise amplitude and a zero mean, delta correlated noise $f_j^{(\nu)}(t)$:
\begin{align}
&F_{j}^{\mathrm{(th)}}(t)=\sqrt{\bar{n}_{j,\mathrm{bath}}+\frac{1}{2}} f^{\mathrm{(th)}}_{j}(t),\nonumber\\
&F_{m}^\mathrm{(in)}(t)=\sqrt{\frac{\eta^{2}}{2}}f^\mathrm{(in)}_{m}(t),\\
&F_{s}^\mathrm{(in)}(t)=\sqrt{\frac{1-\eta^{4}}{2}}f^\mathrm{(in)}_{s}(t),
\end{align}
where $\eta^2 \approx 0.8$ is the power transmission coefficient of the light between the spin and the membrane and $\bar{n}_{j,\mathrm{bath}}$ is the number of thermal phonons in the individual system. The thermal noise amplitude is calculated from the fluctuation dissipation theorem while for the derivation of the backaction noise we refer to \cite{karg2019}. The number of thermal phonons of the membrane $\bar{n}_{m,\mathrm{bath}}$ was measured by homodyne detection in presence of all laser beams but without loading the atoms. This calibrated value agrees very well with an estimation from comparing the spectral linewidth in presence of the cooling and coupling beams with the spectral linewidth of the uncooled membrane and the calculated room temperature occupation of the membrane. We assumed the spin pumping to be perfect such that the spin oscillator environment is in its quantum mechanical ground state (i.e. $\bar{n}_{s,\mathrm{bath}} = 0$).

The approach given in \cite{norrelykke2011} allows for an exact simulation of the stochastic dynamics for a single oscillator for arbitrary time steps, which we extend to the case of two coupled oscillators with delay. This is done by calculating for each time step the coherent evolution and the noise separately:
\begin{equation}
\begin{pmatrix}
\tilde{X}_{m}(t_{i+1}) \\
\tilde{P}_{m}(t_{i+1}) \\
\tilde{X}_{s}(t_{i+1}) \\
\tilde{P}_{s}(t_{i+1})
\end{pmatrix}=\mathrm{e}^{-\mathbf{M} \Delta t}\begin{pmatrix}
\tilde{X}_{m}(t_i) \\
\tilde{P}_{m}(t_i) \\
\tilde{X}_{s}(t_i) \\
\tilde{P}_{s}(t_i)
\end{pmatrix}+\begin{pmatrix}
\Delta \tilde{X}_{m}^{t_i \rightarrow t_{i+1}} \\
\Delta \tilde{P}_{m}^{t_i \rightarrow t_{i+1}} \\
\Delta \tilde{X}_{s}^{t_i \rightarrow t_{i+1}} \\
\Delta \tilde{P}_{s}^{t_i \rightarrow t_{i+1}}
\end{pmatrix},
\end{equation}
where $\Delta t = t_{i+1} - t_i$ is one simulation time step, and $\Delta \tilde{X}_{j}^{t_i \rightarrow t_{i+1}}$, $\Delta \tilde{P}_{j}^{t_i \rightarrow t_{i+1}}$ are terms for the stochastic noise which enters the system in between time $t_i$ and $t_{i+1}$. We performed the simulation at time steps comparable to the oscillation period $\Omega_{m}^{-1}$. Thus, the noise terms $\Delta \tilde{X}_{j}^{t_i \rightarrow t_{i+1}}$ and $\Delta \tilde{P}_{j}^{t_i \rightarrow t_{i+1}}$ are correlated which is taken into account by following the calculation of noise variances and covariances in \cite{norrelykke2011}. Because the coupling between the two oscillators is much slower than the simulation time step $g \ll \Omega_{m} \approx \Delta t^{-1}$ we neglect the correlation of noise building up between the oscillators during one simulation step. Thus, we can treat the noise of both oscillators separately. In order to simulate the system more efficiently, we perform the simulation in time steps of multiples of one frame rotation $\Delta t = k\cdot 2\pi/\Omega_m$, $k=1,2,3...$ such that the noise amplitudes [proportional to $\sin(\Omega_m t), \cos(\Omega_m t)$, see Eq.~\eqref{1:EOM}] are the same for each step of the simulation.

\subsection{Fit function for the power spectral density of mechanical displacement}\label{sec:PSD coupled-mode}
In this section, we provide some details on the coupled-mode model used for fitting the power spectral density of the mechanical displacement shown in Fig.~\ref{fig: continuous pump 2}(b). For this, we first Fourier transform the equations of motion Eqs.~\eqref{eq.equations of motion 1} and \eqref{eq.equations of motion 2}, which allows us to derive the following effective susceptibilities
\begin{align}
&\chi_{m, 0}(\omega)^{-1} X_{m}(\omega)+2 g\, \mathrm{e}^{\mathrm{i} \omega \tau} X_{s}(\omega) =-\sqrt{2 \gamma_{m}} F_{m}^{(\mathrm{tot})}(\omega), \\
&\chi_{s, 0}(\omega)^{-1} X_{s}(\omega)+2 g\,\mathrm{e}^{\mathrm{i} \omega \tau} X_{m}(\omega) =-\sqrt{2 \gamma_{s}} F_{s}^{(\mathrm{tot})}(\omega),
\end{align}
where we have defined the individual oscillator susceptiblities as
\begin{equation}
\chi_{i, 0}(\omega)=\frac{\Omega_{i}}{\Omega_{i}^{2}-\omega^{2}-\mathrm{i} \omega \gamma_{i}}.
\end{equation}
Solving for $X_{m}$ and $X_{s}$ yields 
\begin{align}
X_{m}(\omega) =&\chi_{m,\mathrm{eff}}(\omega)\big[-\sqrt{2 \gamma_{m}} F_{m}^{(\mathrm{tot})}(\omega)\nonumber\\
&+ 2 g\, \mathrm{e}^{\mathrm{i} \omega \tau} \sqrt{2 \gamma_{s}} \chi_{s, 0}(\omega) F_{s}^{(\mathrm{tot})}(\omega)\big],
\end{align}
\begin{align}
X_{s}(\omega) =&\chi_{s, \mathrm{eff}}(\omega)\big[-\sqrt{2 \gamma_{s}} F_{s}^{(\mathrm{tot})}(\omega)\nonumber\\
&+2 g\, \mathrm{e}^{\mathrm{i} \omega \tau} \sqrt{2 \gamma_{m}} \chi_{m, 0}(\omega) F_{m}^{(\mathrm{tot})}(\omega)\big],
\label{eq:bare_susceptibility}
\end{align}
where we have introduced the effective susceptibilities of the membrane and spin oscillators as
\begin{align}
&\chi_{m,\mathrm{eff}}(\omega)^{-1} =\chi_{m, 0}(\omega)^{-1}-4 g^{2} \mathrm{e}^{\mathrm{i} 2 \omega \tau} \chi_{s, 0}(\omega), \label{eq:membrane_susceptibility}\\
&\chi_{s,\mathrm{eff}}(\omega)^{-1} =\chi_{s, 0}(\omega)^{-1}-4 g^{2} \mathrm{e}^{\mathrm{i} 2 \omega \tau} \chi_{m, 0}(\omega).
\end{align}
We used this model to fit the power spectral densities of the mechanical displacement spectra [see Fig.~\ref{fig: continuous pump 2}(b)] using as fit function $a^{2}|\chi_{m,\mathrm{eff}}(\omega)|^{2}$ where $a$ is a global scaling factor accounting for the noise terms driving the system. 

\subsection{Derivation of the sympathetic cooling rate}\label{sec:sympathetic cooling}
Here, we derive the sympathetic cooling rate for the mechanical oscillator given in Eq.~\eqref{eq:delay_gamma_s}. For this, let us first write Eq.~\eqref{eq:membrane_susceptibility} explicitly
\begin{align}
\chi_{m,\mathrm{eff}}(\omega)^{-1}=&\frac{1}{\Omega_{m}}\Bigg(\Omega_{m}^{2}-\omega^{2}-\mathrm{i} \omega \gamma_{m}\nonumber\\
&-4 g^{2} \mathrm{e}^{\mathrm{i} 2 \omega \tau}  \frac{\Omega_{m} \Omega_{s}\left(\Omega_{s}^{2}-\omega^{2}+\mathrm{i}  \omega \gamma_{s}\right)}{\left(\Omega_{s}^{2}-\omega^{2}\right)^{2}+\left(\omega \gamma_{s}\right)^{2}}\Bigg),
\end{align}
which can be written in the form of
\begin{equation}
\chi_{m,\mathrm{eff}}(\omega)^{-1}=\frac{1}{\Omega_{m}}\left[\Omega_{m}^{2}-\delta\Omega_{\rm{shift}}^{2}-\omega^{2}-\mathrm{i} \omega\left(\gamma_{m}+\gamma_\mathrm{sym}\right)\right].
\label{eq:standard_form_eff_susceptibility}
\end{equation}
Here, we have defined an effective frequency shift $\delta\Omega_\mathrm{shift}$ and the sympathetic cooling rate $\gamma_\mathrm{sym}$, which for $\omega=\Omega_m$ read
\begin{align}
\delta\Omega_{\mathrm{shift}}^{2} = &\frac{4 g^{2} \Omega_{m} \Omega_{s}}{\left(\Omega_{s}^{2}-\Omega_{m}^{2}\right)^{2}+\left(\Omega_{m} \gamma_{s}\right)^{2}}\nonumber\\
&\times\big[\left(\Omega_{s}^{2}-\Omega_{m}^{2}\right) \cos \left(2 \Omega_{m} \tau\right)
-\Omega_{m} \gamma_{s} \sin \left(2 \Omega_{m} \tau\right)\big],
\end{align}
and
\begin{align}
\gamma_\mathrm{sym} = &\frac{4 g^{2} \Omega_{m} \Omega_{s}}{\left(\Omega_{s}^{2}-\Omega_{m}^{2}\right)^{2}+\left(\Omega_{m} \gamma_{s}\right)^{2}}\nonumber\\
&\times \left[\gamma_{s} \cos \left(2 \Omega_{m} \tau\right)+\frac{\Omega_{s}^{2}-\Omega_{m}^{2}}{\Omega_{m}} \sin \left(2 \Omega_{m} \tau\right)\right].
\end{align}
For $\Omega_{s} \approx \Omega_{m}$ and large spin damping $\gamma_{s} > g$, we get a simplified expression for the frequency shift and sympathetic cooling rate [Eq.~\eqref{eq:delay_gamma_s}]
\begin{align}
&\delta\Omega_{\text {shift}}^{2} \approx \frac{4 g^{2}\Omega_{m}}{4 \delta^{2}+\gamma_{s}^{2}}\left[2 \delta \cos \left(2 \Omega_{m} \tau\right) - \gamma_{s} \sin \left(2 \Omega_{m} \tau\right)\right], \\
&\gamma_{\mathrm{sym}} \approx \frac{4 g^{2}}{4 \delta^{2}+\gamma_{s}^{2}}\left[\gamma_{s} \cos \left(2 \Omega_{m} \tau\right)+2 \delta \sin \left(2 \Omega_{m} \tau\right)\right],
\end{align}
where $\delta = \Omega_{s} - \Omega_{m}$.

\subsection{Routh-Hurwitz stability criterion of the coupled system}\label{sec:Hurwitz}

\begin{figure}[t]
\centering
\includegraphics{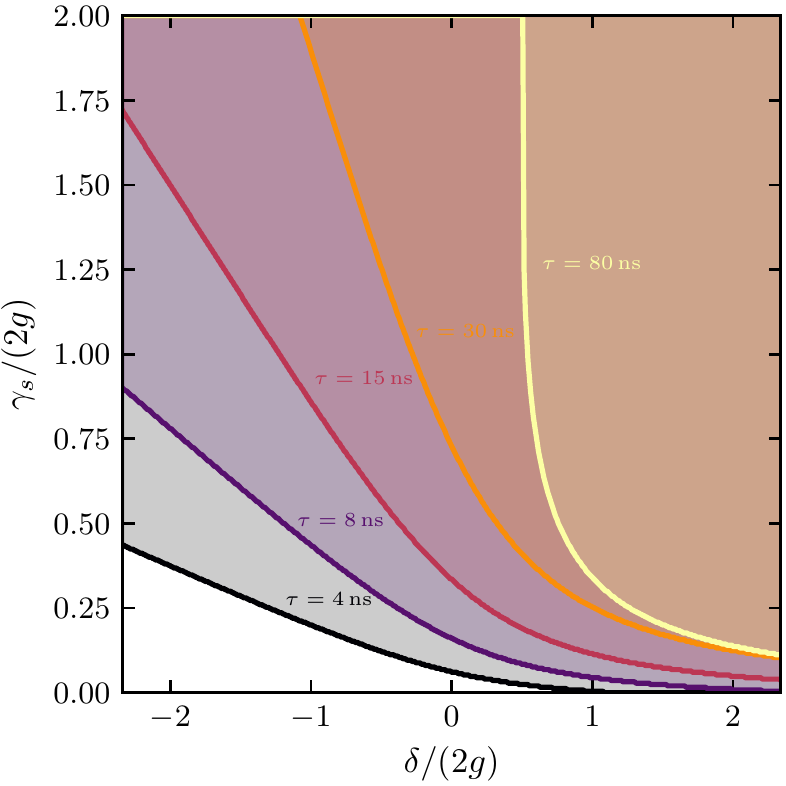}
\caption{Evaluation of the stability of the coupled system using the Routh-Hurwitz criterion: The colored regions (i.e. region above each solid line) show the sets of parameters for which the coupled dynamics is stable for a given value of the feedback delay. Without propagation delay, every set of detunings and spin damping leads to stable dynamics. For $\tau = \SI{80}{\nano\second}$ we have $\Omega_m\tau \approx 1$ thus the validity of the Taylor expansion of the exponential function in presence of small delays reaches its limit. For the stability estimations shown here we used $2 g = 2\pi\times \SI{6.8}{\kilo\hertz}$, $\gamma_m = 2\pi\times \SI{262}{\hertz}$, and $\Omega_m = 2\pi\times \SI{1.957}{\mega\hertz}$}
\label{fig:Hurwitz}
\end{figure}

In this section we present a stability analysis in which the Routh-Hurwitz criterion \cite{hofer2008} from control theory is applied to our linearly coupled spin-membrane oscillators. The criterion provides a convenient means to assess the stability of our linear systems without solving the equations of motion. In this treatment, we exclude the Langevin noise, as we are interested to see if the delayed coupled oscillator dynamics is stable by itself. We then explore the experimental parameter space to see under which conditions the coupled system becomes unstable. We take the equations of motion for the delayed coupled system  Eqs.~\eqref{eq.equations of motion 1} and \eqref{eq.equations of motion 2} neglecting the noise terms
\begin{align}
&\ddot{X}_{m}+\gamma_{m} \dot{X}_{m}+\Omega_{m}^{2} X_{m} =-2 g \Omega_{m} X_{s}(t-\tau), \label{eq:Hurwitz,eom1}\\
&\ddot{X}_{s}+\gamma_{s} \dot{X}_{s}+\Omega_{s}^{2} X_{s} =-2 g \Omega_{s} X_{m}(t-\tau).
\label{eq:Hurwitz,eom2}
\end{align}
Substituting the ansatz $X_{j}(t) = X_{{j}}(s)\, \mathrm{e}^{s t}$ where $s\in\mathbb{C}$ yields
\begin{align}
&\left(s^{2}+s \gamma_{m}+\Omega_{m}^{2}\right) X_{m}(s) =-2 g \Omega_{m} \,\mathrm{e}^{-s \tau} X_{s} (s), \label{eq:sdomain1}\\
&\left(s^{2}+s \gamma_{s}+\Omega_{s}^{2}\right) X_{s}(s) = -2 g \Omega_{s} \,\mathrm{e}^{-s \tau} X_{m}(s).
\label{eq:sdomain2}
\end{align}
Solving the simultaneous equations Eqs.~\eqref{eq:sdomain1} and \eqref{eq:sdomain2}, we obtain the characteristic equation for non-trivial solutions $X_{m} \neq 0$, 
\begin{equation}
\left(s^{2}+s \gamma_{m}+\Omega_{m}^{2}\right)\left(s^{2}+s \gamma_{s}+\Omega_{s}^{2}\right) - 4 g^{2} \Omega_{m} \Omega_{s} \,\mathrm{e}^{-2 s \tau} =0.
\end{equation}
For clarity, we consider here small propagation delays $\tau \ll {1}/{\Omega_{j}}$ and apply a first order Taylor expansion $\exp({-2s\tau})\approx (1-2s\tau)$ (in the actual simulation we keep terms up to 4th order). We then obtain
\begin{align}
0=& s^{4} + (\gamma_{s} + \gamma_{m}) s^{3} + (\Omega_{m}^{2} + \Omega_{s}^{2} + \gamma_{m} \gamma_{s}) s^{2} \nonumber\\
 &+ ( \Omega_{s}^{2} \gamma_{m} + \Omega_{m}^{2} \gamma_{s} + 8 g^{2} \Omega_{m} \Omega_{s} \tau) s \nonumber \\
 &+ \Omega_{m} \Omega_{s} (\Omega_{m} \Omega_{s} - 4 g^{2}).
\label{eq:polynomial}
\end{align}
Having our dynamics in this polynomial form, we can define the polynomial coefficients of a fourth order polynomial by
\begin{equation}
p(s) = a_{4} s^{4}+a_{3} s^{3}+a_{2} s^{2}+a_{1} s+a_{0}=0, \: a_4>0.
\end{equation}
\\
In order to apply the Routh-Hurwitz criterion, the so-called Hurwitz matrix containing the polynomial coefficients has to be defined. For a fourth order polynomial this matrix reads
\begin{equation}
H_{4}=\begin{pmatrix}
a_{3} & a_{1} & 0 & 0 \\
a_{4} & a_{2} & a_{0} & 0 \\
0 & a_{3} & a_{1} & 0 \\
0 & a_{4} & a_{2} & a_{0}
\end{pmatrix}.
\end{equation}
According to the Routh-Hurwitz criterion, the system dynamics is asymptotically stable if all the principal minors of the Hurwitz matrix are non-zero and positive. Application of the Hurwitz criterion leads to the following stability criteria for a fourth order polynomial system:
\begin{align}
&\Delta_{1}=\left|a_{3}\right|>0,\\
&\Delta_{2}=\begin{vmatrix}
a_{3} & a_{1} \\
a_{4} & a_{2}
\end{vmatrix}=a_{2} a_{3}-a_{4} a_{1}>0, \\
&\Delta_{3}=\begin{vmatrix}
a_{3} & a_{1} & 0 \\
a_{4} & a_{2} & a_{0} \\
0 & a_{3} & a_{1}
\end{vmatrix}=a_{1} \Delta_{2}-a^{2}_{3} a_{0}>0, \\
&\Delta_{4}=\det(H_4) = a_0\cdot \Delta_3 > 0.
\end{align}
In our system, the coefficients are given explicitly by
\begin{align}
&a_{4} =1, \\
&a_{3} =\gamma_{s}+\gamma_{m}, \\
&a_{2} =\Omega_{s}^{2}+\Omega_{m}^{2}+\gamma_{s} \gamma_{m}, \\
&a_{1} =\gamma_{m} \Omega_{s}^{2}+\gamma_{s} \Omega_{m}^{2} + 8g^{2} \Omega_{m} \Omega_{s} \tau, \\
&a_{0} =\Omega_{s} \Omega_{m}\left(\Omega_{s} \Omega_{m} - 4 g^{2} \right).
\end{align}
Since $\Omega_{s} \Omega_{m} \gg 4g^{2}$, all coefficients are positive. Thus, the criterion $\Delta_1$ is fulfilled and the criterion $\Delta_4$ depends directly on the criterion $\Delta_3$. Therefore, only $\Delta_{2}$ and $\Delta_{3}$ are left to be checked. In order to get an intuition on the stability for different parameters, Fig.~\ref{fig:Hurwitz} shows the stable regions as a function of spin damping, detuning and delay.

\vfill

\bibliographystyle{apsrev}

\end{document}